\documentclass{PoS}

\title{Chiral behaviour of the pion decay constant \\ in $N_\mathrm{f}=2$ QCD}

\newcommand{\preprintline}{\vspace{3cm}\newline
\rightline{\parbox{2.9cm}{\large\tt DESY 13-213}}
}

\ShortTitle{$f_\pi$ chiral behaviour in $N_\mathrm{f}=2$ QCD}

\author{ 
	\speaker{Stefano Lottini} \qquad \raisebox{0.9cm}{\hspace{-3.6cm}\includegraphics[width=1.8cm,angle=0]{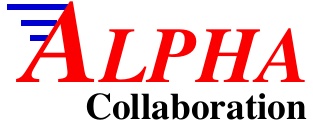}}
		\\
		NIC, DESY -- Platanenallee 6, 15738 Zeuthen, Germany\\
		E-mail: \email{stefano.lottini@desy.de} \\
}

\author{
		\vspace{-0.8cm}
		\textbf{for the ALPHA Collaboration}
}

\abstract{
 As increased statistics and new ensembles with light pions have
 become available within the CLS effort, we complete previous
 work by inspecting the chiral behaviour of the pion decay constant.
 We discuss the validity of Chiral Perturbation Theory ($\chi$PT) and
 examine the results concerning the pion decay constant and the ensuing
 scale setting, the pion mass squared in units of the quark mass,
 and the ratio of decay constants $f_\mathrm{K}/f_\pi$; along
 the way, the relevant low-energy constants of SU(2) $\chi$PT are estimated.
 All simulations were performed with two dynamical flavours of
 nonperturbatively O(a)-improved Wilson fermions, on volumes with
 $m_\pi L \geq 4$, pion masses \mbox{$\geq$ 192 MeV} and lattice spacings
 down to 0.048 fm. Our error analysis takes into account
 the effect of slow modes on the autocorrelations.
 \preprintline
}

\FullConference{31st International Symposium on Lattice Field Theory LATTICE 2013\\
July 29 -- August 3, 2013\\
Mainz, Germany}

\newcommand{\eq}{\begin{equation}}
\newcommand{\qe}{\end{equation}}

\newcommand{\bei}{\begin{itemize}}
\newcommand{\eei}{\end{itemize}}
\newcommand{\bea}{\begin{eqnarray*}}
\newcommand{\eea}{\end{eqnarray*}}

\newcommand{\yt}{y}

\newcommand{\lbar}{{\overline{\ell}}}

\begin{document}

\section{Introduction}
The nonperturbative features of quantum chromodynamics (QCD)
are more and more prominent as one approaches the low-energy regime;
there, the most successful strategy to determine its properties is,
to date, numerical simulation on the lattice, complemented by insight
provided by chiral perturbation theory ($\chi$PT) as to how
the chiral limit is reached.

In the past years, new $N_\mathrm{f}=2$ ensembles have been produced
within the Coordinated Lattice Simulations (CLS) effort,
closer and closer to both the physical point and
the continuum limit. Here we report on the chiral behaviour of
the pion decay constant $f_\pi$ (and related quantities)
obtained from the latest set of CLS ensembles,
and the subsequent determinations of the $SU(2)$ $\chi$PT
low-energy constants involved.
The analysis presented here can be seen as a complement
to Ref.~\cite{Fritzsch:2012wq}
(to which we refer for most of the setup details),
where analogous investigations were
carried on, among other topics,
concerning the kaon decay constant $f_\mathrm{K}$.

All ensembles were generated using $O(a)$-improved
Wilson lattice action with two degenerate dynamical flavours,
implementing either Domain-Decomposition \cite{Luscher:2005rx}
or Mass-Preconditioning \cite{Marinkovic:2010eg}.
Table \ref{table:ensembles} summarises the relevant information
on the ensembles employed in the present analysis.

Two-point functions
were measured
using 10 to 20 $U(1)$ stochastic sources per configuration,
and subsequently
used to extract meson- and PCAC-masses and decay constants,
as detailed in \cite{Fritzsch:2012wq},
with a statistical uncertainty on
the level of 1\% or better.
For renormalisation,
the $b$-factors come from one-loop perturbation theory
\cite{Sint:1997dj}, and the $Z$-factors
are nonperturbatively determined \cite{Fritzsch:2012wq}.

\section{Chiral analyses}
In the data analysis, first an independent variable is built,
parameterising the approach to the chiral and the physical points;
in contrast to the one employed in \cite{Fritzsch:2012wq}, here we
deal with the purely pionic (i.e.~light-light flavours) variable
\eq
	\yt_1 = \frac{m_\pi^2(m_q)}{8\pi^2f_\pi^2(m_q)}\;\;,
\qe
appearing as expansion parameter in the $\chi$PT formulae.
$m_\pi$ and $f_\pi$ are the measured
quantities at the finite quark masses;
the physical point corresponds to $\yt_\pi\simeq0.01353$.

Generally, fits are performed, simultaneously for all values of $\beta$,
to the functional form coming from $\chi$PT,
taking correlations among measurement into account.
Our goal being the chiral and physical
points, we inspect fits in the
range $m_\pi \leq m_\pi^\mathrm{(cut)}$ with 
$m_\pi^\mathrm{(cut)} \leq 650, 500, 390, 345$ MeV;
as it turns out, beyond NLO, a quadratic term
has to be included for the fit to be stable
in $m_\pi^\mathrm{(cut)}$; systematic uncertainties in the
resulting parameters were estimated also by altering the fit
functions.

\begin{table}
\begin{center}
\begin{tabular}{||rcrr||rcrr||rcrr||}
\hline
	\multicolumn{4}{||c||}{$\beta=5.2$~~~($a\simeq 0.075$~fm)} &
	\multicolumn{4}{  c||}{$\beta=5.3$~~~($a\simeq 0.065$~fm)} &
	\multicolumn{4}{  c||}{$\beta=5.5$~~~($a\simeq 0.048$~fm)} \\
\hline
	& $m_\pi$ & MDU & $\frac{\mathrm{MDU}}{\tau_\mathrm{exp}}$ &
	& $m_\pi$ & MDU & $\frac{\mathrm{MDU}}{\tau_\mathrm{exp}}$ &
	& $m_\pi$ & MDU & $\frac{\mathrm{MDU}}{\tau_\mathrm{exp}}$ \\
\hline
A2 & 629 & 8000 & 120 & E4  & 580 &  2496 &  10 & N4 & 551 & 3752 &   4 \\
A3 & 492 & 8032 & 120 & E5f & 436 & 16000 &  60 & N5 & 440 & 3808 &   4 \\
A4 & 383 & 8096 & 120 & E5g & 436 & 16000 & 120 & N6 & 340 & 8040 &  40 \\
A5 & 330 & 4004 & 160 & F6  & 311 &  4800 &  36 & O7 & 267 & 3920 &  20 \\
B6 & 281 & 1272 &  50 & F7  & 266 &  9416 &  70 &    &     &      &     \\
   &     &      &     & G8  & 192 &  1114 &  20 &    &     &      &     \\
\hline
\end{tabular}
\caption{Overview of the ensembles used.
For the three bare lattice couplings $\beta$,
three quantities are shown: approximate pion mass (expressed in MeV),
ensemble extent in Molecular Dynamic Units (MDU), and the latter in units of
the autocorrelation time associated to the slowly-decaying modes
(see text).
For each ensemble 10 stochastic sources were used to extract the
two-point functions, except for the last entry
of each $\beta$ (20 sources).
All ensembles have $m_\pi L\geq 4$ and time extent $T=2 L$.
The two `E5' ensembles differ in the trajectory length (respectively $\tau=0.5$ and $2.0$~MDU).}
\label{table:ensembles}
\end{center}
\end{table}

Correlations are propagated down to the
final quantities. Special care is taken for the 
integrated autocorrelation time, by applying the technique developed in
\cite{Schaefer:2010hu}: the effects of slow modes in the transition matrix,
to which an observable can couple, are estimated by attaching
an $\exp(-t/\tau_\mathrm{exp})$ tail, with $\tau_\mathrm{exp}$ coming from prior
determinations,
to the autocorrelation function where its statistical error is too large
for direct determination.
For all quantities, however, the largest contribution (about 30 to 60\%)
to the full uncertainty comes from the error on the renormalisation factor $Z_A$.

\subsection{Analysis of $f_\pi$}
Here and in the following, we denote decay constants in lattice units with an
uppercase symbol, as in $F_\pi(\beta)=a(\beta)f_\pi$, with $f_\pi$ taking the
value $f_\pi^\mathrm{phys}=130.4$~MeV at the physical point.
The outcome of this study, focused on how and to which extent can $\chi$PT
be trusted to describe the behaviour of $f_\pi$, will then provide a
scale-setting prescription.

The main fits are performed to the $\chi$PT NLO functional \cite{Gasser:1983yg},
plus a quadratic term:
\eq
	F_\pi(\yt_1)=F_\pi^\mathrm{phys}(\beta)
		\{
			1+\alpha_4(\yt_1-\yt_\pi)-\yt_1\log\yt_1+\yt_\pi\log\yt_\pi+B(\yt_1-\yt_\pi)^2
		\}\;\;,
	\label{eq:fpi_nnlo_fit}
\qe
where the three $F_\pi$ (one for each $\beta$) will yield the scale setting and $\alpha_4$ encodes the
$SU(2)$ NLO low-energy constant (LEC) $\lbar_4$:
\eq
	\lbar_4=\log\frac{\Lambda_4^2}{\mu^2}\Big|_{\mu\leftarrow m_\pi^\mathrm{phys}=134.8~\mathrm{MeV}}
	\quad;\qquad
	\lbar_4=\frac{\alpha_4}{1-\alpha_4\yt_\pi+\yt_\pi\log\yt_\pi}-\log\yt_\pi\;\;.
\qe

\begin{figure}
\begin{center}
\begin{minipage}{0.535\textwidth}
	\hspace{-1.2cm}\includegraphics[width=1.2\textwidth]{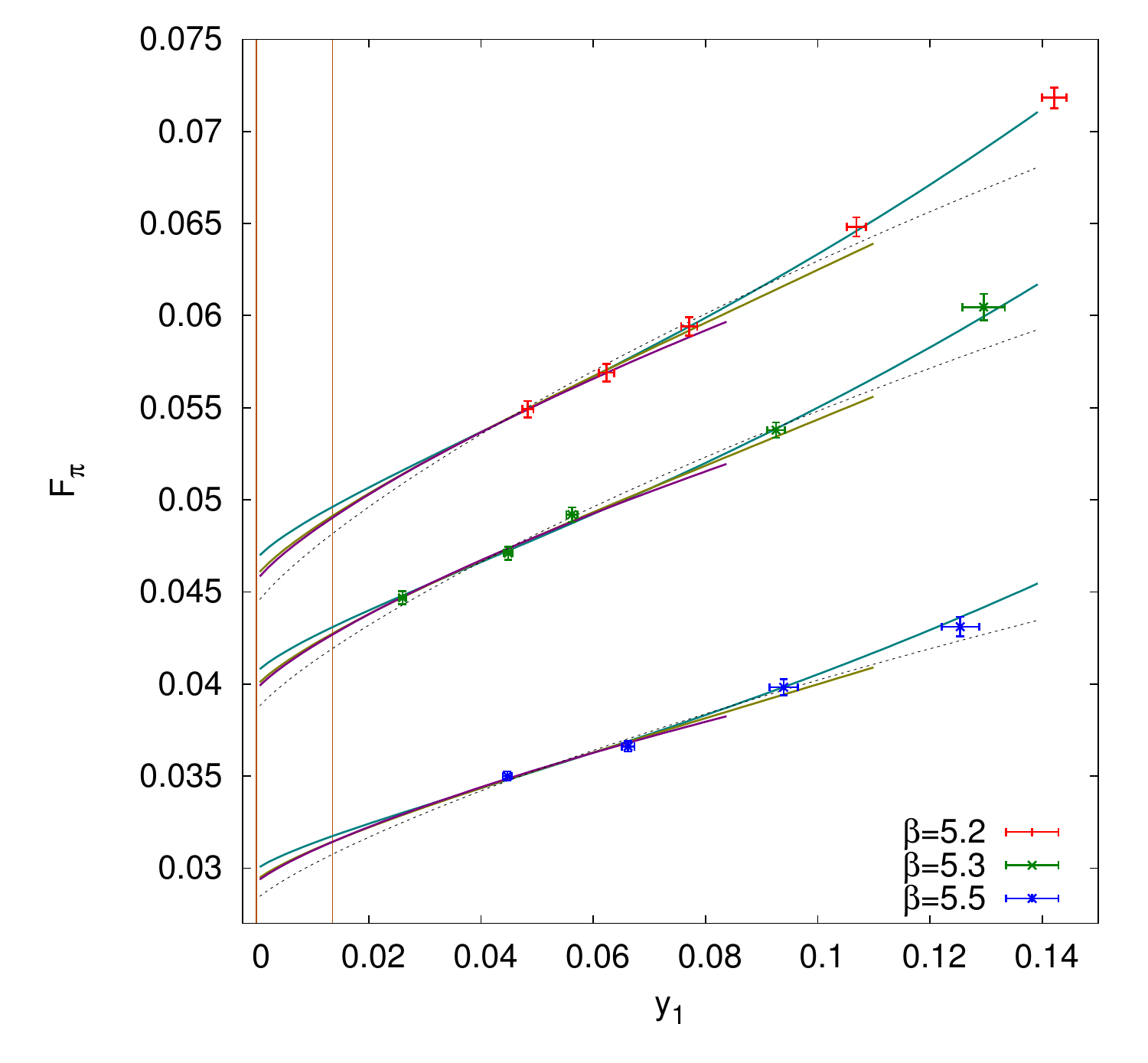}
\end{minipage}\hspace{0.5cm}\begin{minipage}{0.42\textwidth}
	\includegraphics[width=0.99\textwidth]{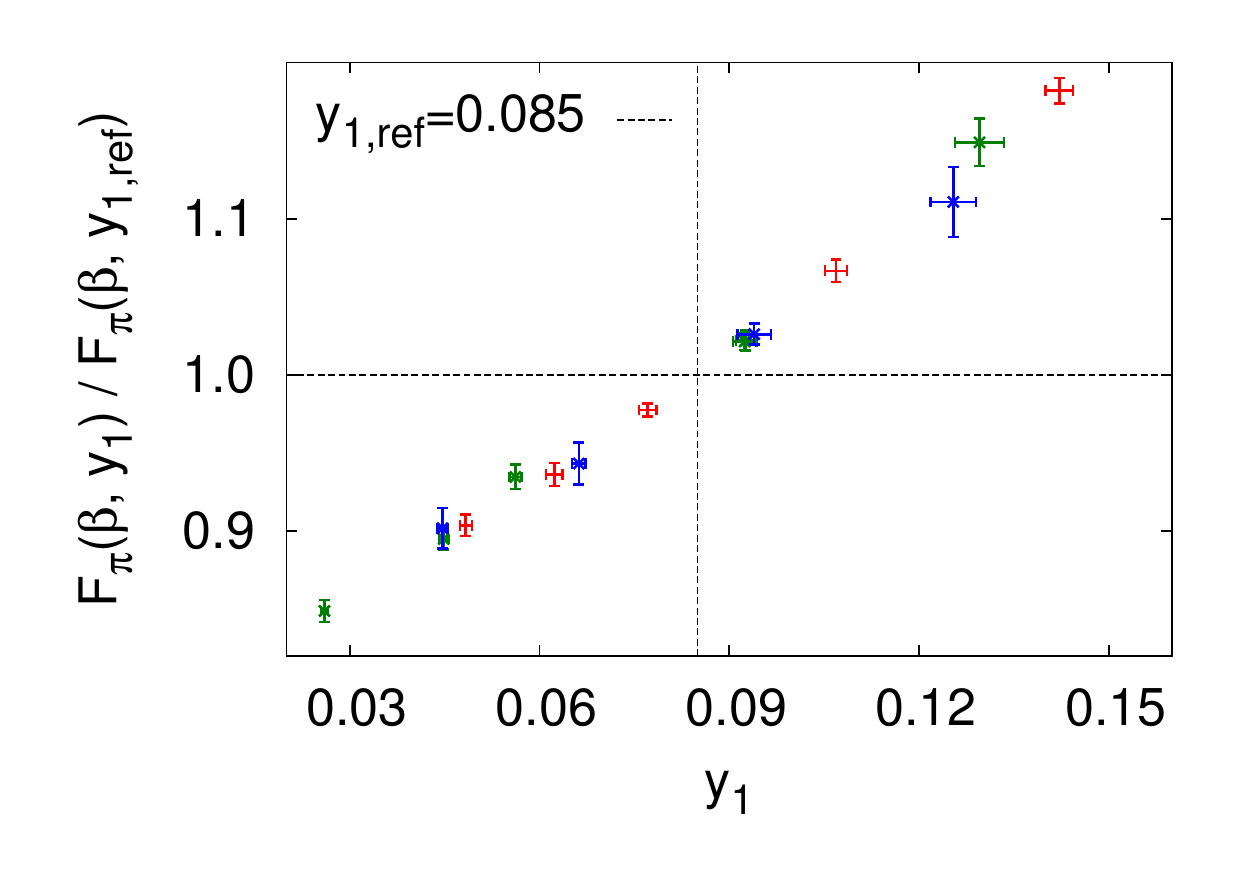}\\
	\includegraphics[width=0.99\textwidth]{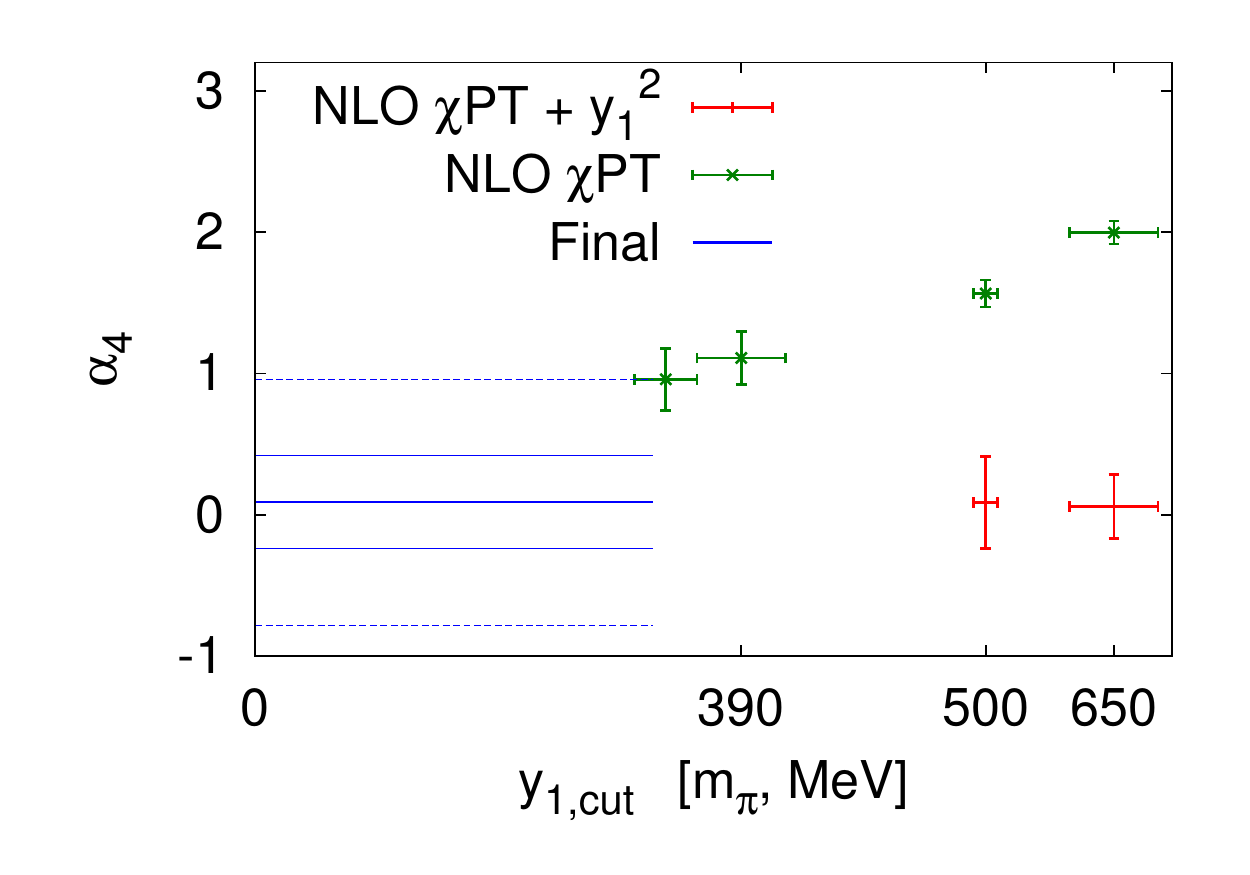}
\end{minipage}
\caption{Behaviour of $F_\pi$.
	\textit{Left:} $F_\pi(\yt_1)$, measured points at the three $\beta$ and fit to Eq.~\protect\ref{eq:fpi_nnlo_fit} for 
	$m_\pi^\mathrm{(cut)}=500, 390, 345$~MeV (solid lines);
	the dashed line is a fit with $B=0$ and $m_\pi^\mathrm{(cut)}=500$~MeV added for illustration.
	The vertical line marks $\yt_\pi$.
	\textit{Right, top:} collapse-plot of the same data,
	rescaled to the corresponding interpolated $F_\pi(\yt_{1,\mathrm{ref}}=0.085)$
	(same colours as in left plot).
	\textit{Right, bottom:} resulting fit parameter $\alpha_4$ for the fit with and without quadratic term for various
	$m_\pi^\mathrm{(cut)}$; horizontal lines mark mean value, one-sigma statistical error band, and systematic uncertainty
	of the final result (equivalent to $\lbar_4$ of Eq.~\protect\ref{eq:ellbar4_result}).
	Note how the \mbox{`NLO $\chi$PT + $y_1^2$'}
	points are stable at the value suggested by the \mbox{$m_\pi^\mathrm{(cut)}\to 0$}
	trend of the `NLO $\chi$PT' points.
}
\label{fig:fpi_fits}
\end{center}
\end{figure}

This functional form fits the data in a stable way for various $m_\pi^\mathrm{(cut)}$
(Fig.~\ref{fig:fpi_fits}, left),
and we take the results from the cut at 500 MeV as median value; systematic uncertainties
come from comparing functional forms without $B$ and/or the chiral logarithms.
The absence of $am_q$-effects is illustrated by the collapse-plot of
Fig.~\ref{fig:fpi_fits}, right top, where data for each $\beta$ are
rescaled to a common curve.
An important remark is the following: the result we get for $B$ is such that, in the
data range, the $\yt_1^2$-term contributes by as much as $\sim$15\% to the curve.
This is reflected in Fig.~\ref{fig:fpi_fits}, right bottom, that illustrates the behaviour
of $\alpha_4$ as a function of the pion-mass cut.

This procedure leads to the following determination of the lattice spacing $a(\beta)$,
compatible with the one of \cite{Fritzsch:2012wq} albeit slightly smaller in central value
(the first error is statistical, the second systematic):
\eq
	a(5.2)=0.0750(9)(10)~\mathrm{fm}\;,\;\;
	a(5.3)=0.0652(6)(7)~\mathrm{fm}\;,\;\;
	a(5.5)=0.0480(5)(5)~\mathrm{fm}\;.
\qe
This is also
compatible with an update of \cite{Fritzsch:2012wq},
based on $f_\mathrm{K}$ and including all new ensembles.

Another relevant quantity is the ratio between $f_\pi$ at the physical- and chiral-points,
which -- to NLO -- encodes directly the LEC $\lbar_4$:
\eq
	\frac{f_\pi}{f} = 1.061(6)(16)
		\quad;\qquad
	\lbar_4 = 4.4(4)(9)\;\;;
		\label{eq:ellbar4_result}
\qe
the former falls within the $N_\mathrm{f}=2$ world average by the 2013 FLAG review \cite{flag2013:pre_publication}.

On a related note, we mention that a continuum-extrapolation of $f_\pi r_0$, done in a similar 
fashion, yields for the hadron parameter $r_0$ a slightly smaller (but still compatible) value than 
presented in \cite{Fritzsch:2012wq}: $r_0 = 0.485(7)(7)$~fm.

\subsection{Analysis of $m_\pi^2/M_R$}
We now turn to the ratio of the squared pion mass to the (renormalised) quark mass $M_R$, expressed,
as the chiral condensate below,
in the $\overline{\mathrm{MS}}$ scheme at a scale $\mu=2$~GeV.
The $\chi$PT NLO expression extends the Gell-Mann-Oakes-Renner (GMOR) relation \cite{GellMann:1968rz} to:
\eq
	\frac{m_\pi^2}{4M_R}\Big|_\mathrm{NLO} =
		\frac{\Sigma_0}{f^2}\big\{ 1 + \alpha_3 \yt_1 + \frac{1}{2}\yt_1\log \yt_1 \big\} \;\;,
\qe
with the linear coefficient $\alpha_3$ encoding the NLO low-energy constant
$\lbar_3\big|_\mathrm{NLO} = -2\alpha_3 -\log \yt_\pi$.

One is primarily interested in extracting an estimate for the chiral condensate $\Sigma_0$
in the continuum limit: with this goal in mind, we encode the expected $a^2$-scaling directly in the overall
amplitude of the above formula, and look for the (third root of the) condensate in units of the physical
pion decay constant. We also allow for the usual NNLO analytic term in the fit function:
\eq
	\frac{m_\pi^2}{4M_R} = \Big[
			(S_0+a^2S_1)\frac{F_{\pi,\mathrm{phys}}^3}{F^2}
		\Big]
		\Big\{
			1+\alpha_3 \yt_1 +\frac{1}{2} \yt_1\log\yt_1 + B \yt_1^2
		\Big\}
	\;\;;\;\;\;
	\sqrt[3]{\Sigma_{0,\mathrm{cont}}} = f_{\pi,\mathrm{phys}}~\sqrt[3]{S_0}\;\;;
	\label{eq:chircond_fit}
\qe
(the validity of the Ansatz of $a^2$-scaling is corroborated by variants of the analysis
in which first a $\beta$-dependent amplitude is obtained, then the continuum limit is taken;
the values of $a^2$ used here are those coming from the $f_\pi$-based scale setting).

Eq.~\ref{eq:chircond_fit} describes well the data, 
which lie approximately flat in $\yt_1$ (Fig.~\ref{fig:chircond_AND_fk_over_fpi}, left),
and the value of $S_0$ is stable against the usual variants of the fit procedure;
$\alpha_3$, on the other hand, is more pion-mass-cut- and fit-function-dependent, 
leading to a large systematic error on $\lbar_3$. From the fit to the above equation, 
with $m_\pi^\mathrm{(cut)}=500$~MeV, we get the
following values:
\eq
	\sqrt[3]{\Sigma_{0,\mathrm{cont}}} = 268.1(2.6)(4.9)~\mathrm{MeV}
		\;\;;\;\;\;
	\lbar_3 = 2.4(0.1)(^{+0.7}_{-1.3})\;\;.
\qe
The former result compares favorably to most recent two-flavours
lattice determinations \cite{Brandt:2013dua,Cichy:2013gja}.
We remark that a more first-principle lattice determination of $\Sigma_0$,
minimally relying on $\chi$PT assumptions, is currently being carried on
\cite{Engel:2013rwa} based on the method proposed in \cite{Giusti:2008vb}
and using the same CLS configurations employed here;
the same approach, in the context of twisted-mass fermions, is pursued in 
Ref.~\cite{Cichy:2013gja}, with an outcome which overshoots slightly what found here
-- by about 1.5 standard deviations) --
once recast as $r_0 \sqrt[3]{\Sigma_0}$
(there, this combination is used to overcome current
discrepancies in the physical determination of $r_0$).

\subsection{The ratio $f_\mathrm{K}/f_\pi$}
\begin{figure}
\begin{center}
\includegraphics[width=0.46\textwidth]{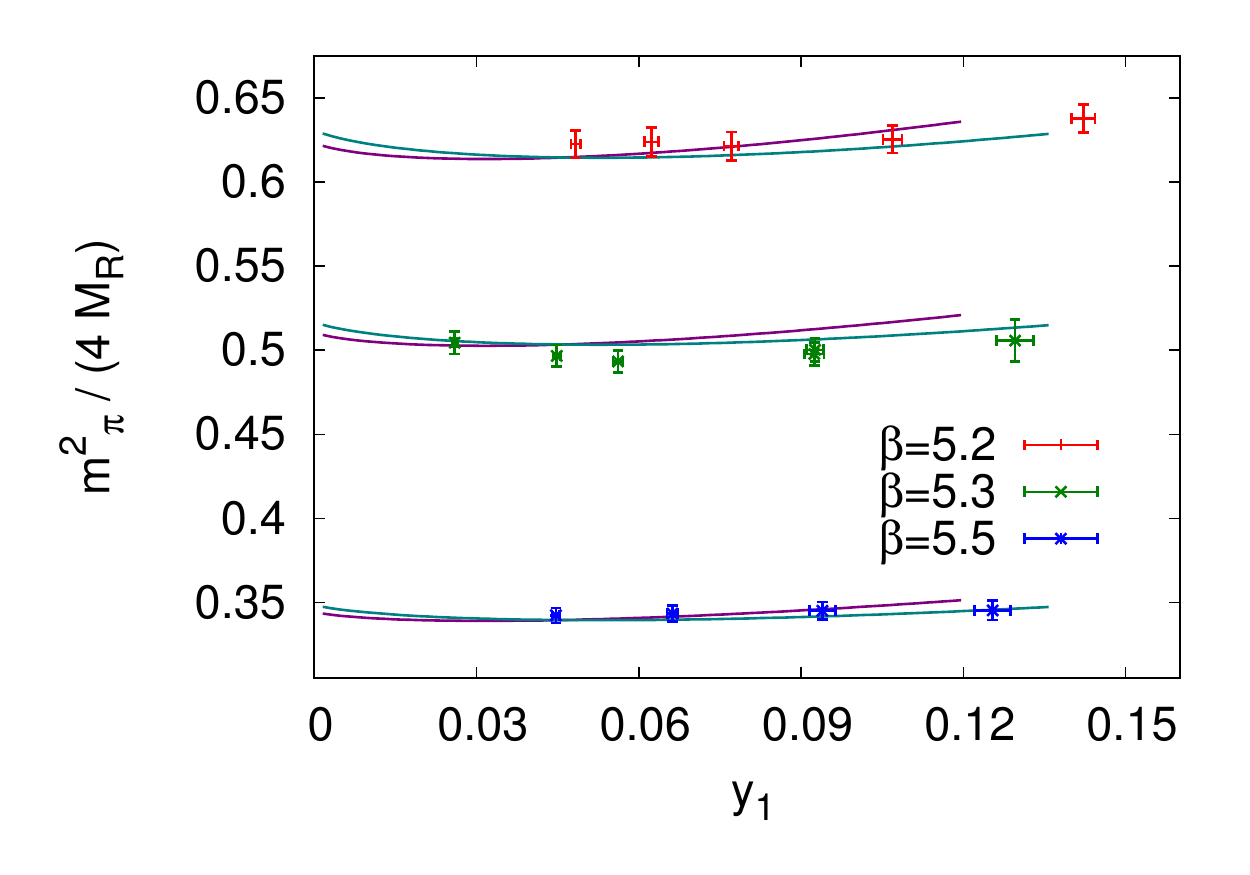}
\includegraphics[width=0.46\textwidth]{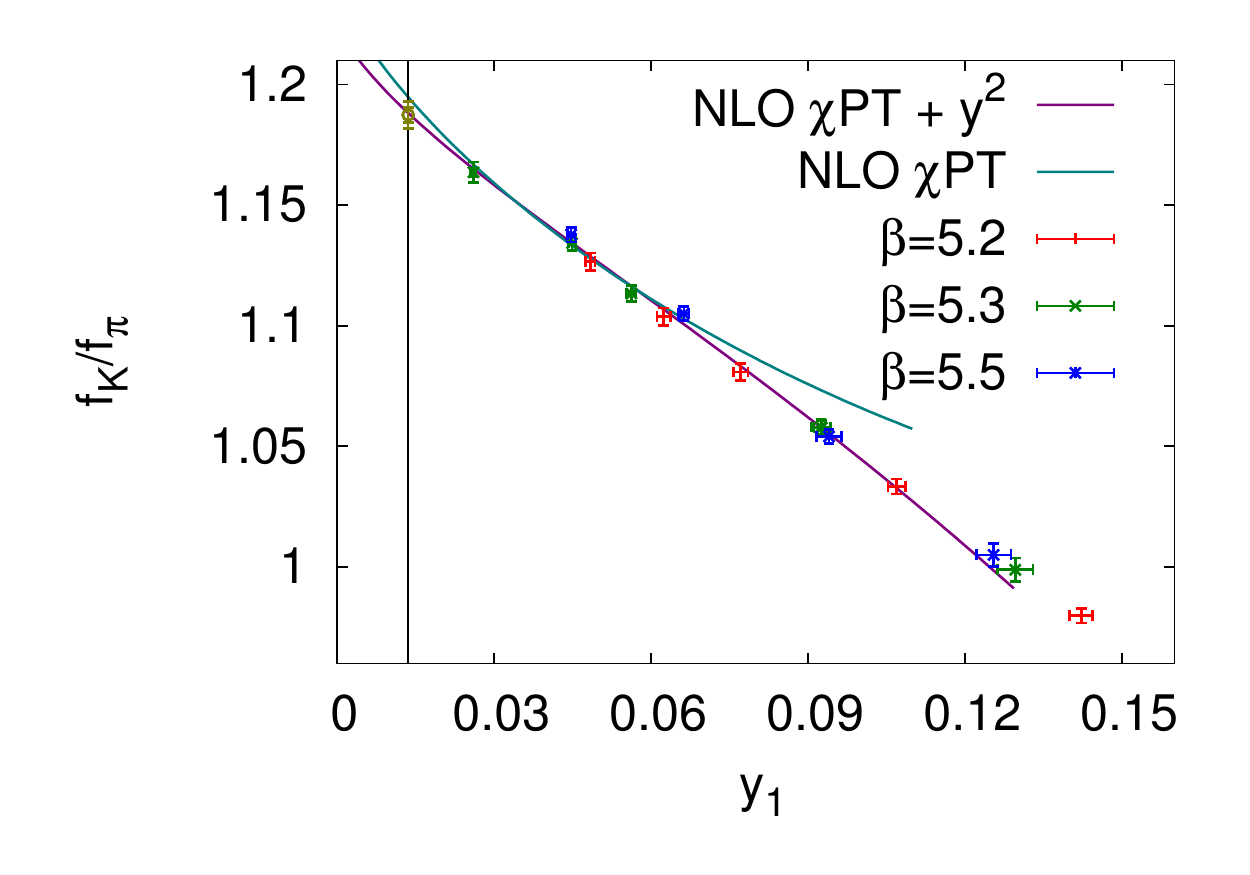}
\caption{\textit{Left:} The quantity $m_\pi^2/(4M_R)$. The solid lines show
	fits to Eq.~\protect\ref{eq:chircond_fit} at two pion-mass cuts (390 and 500 MeV).
	\textit{Right:} Ratio $f_\mathrm{K}/f_\pi$ as a function of $\yt_1$;
	the vertical line marks $\yt_\pi$ and the solid lines are best-fit curves with and
	without the $c_2\yt_1^2$ term
	(and $m_\pi^\mathrm{(cut)}=$ 500 and 345~MeV respectively).}
\label{fig:chircond_AND_fk_over_fpi}
\end{center}
\end{figure}

We now consider a heavier valence quark
and turn to study ``kaon'' properties, by moving along the trajectory
(dubbed `strategy 1' in \cite{Fritzsch:2012wq})
defined by ${m_K}/{f_\mathrm{K}} = {m_K}/{f_\mathrm{K}}|_\mathrm{phys} \simeq 3.19$.
Partially-quenched $\chi$PT dictates the behaviour
of the quantity $f_\mathrm{K}/f_\pi$ \cite{Sharpe:1997by},
which in our range is almost linear in $\yt_1$.
Nevertheless, as usual, we allow for a $\yt_1^2$ term in the fit function:
\eq
	\frac{f_\mathrm{K}}{f_\pi}\Big|_{\mathrm{NLO+}\yt_1^2}  =  R\big\{
			1+c_1 \yt_1 +
			\frac{\yt_1}{2}\log \yt_1
			-\frac{\yt_1}{8}\log\Big( 2\frac{\yt_K}{\yt_1}-1 \Big)
			+c_2 \yt_1^2
		\big\}\;\;;\quad \yt_K = \frac{m_K^2}{8\pi^2 f_\pi^2}\simeq0.182\;.
\qe
Thanks to the stabilising effect of the ensemble with the lightest pion (G8),
the ratio at the physical point can be accurately determined,
see Fig.~\ref{fig:chircond_AND_fk_over_fpi}, right;
moreover, the presence of the quadratic term has an effect similar to that of
Fig.~\ref{fig:fpi_fits}, bottom right. Adding $(af_\pi)^2$-terms as a check did not 
alter the picture at all.
We then quote the result from the fit to the above equation with \mbox{$m_\pi^\mathrm{(cut)}=500$~MeV},
which agrees with most two-flavours lattice estimates, and the CKM matrix element
obtained from the latter through the values in \cite{Antonelli:2010yf, Hardy:2008gy}:
\eq
	\frac{f_\mathrm{K}}{f_\pi}\Big|_\mathrm{phys} = 1.1874(57)(30)\;\;,\;\;\;|V_{us}|=0.2263(13)\;\;.
\qe

\section{Conclusions}
This work presents an analysis of the chiral behaviour of pion-related quantities
based on two-flavour lattice QCD. The agreement of the lattice spacing determination
with the previous ones (from $f_\mathrm{K}$) encourages us; compared to the kaon-based analysis,
however, here $m_\pi^4$-terms are necessary to stabilise the fits.
The ubiquitous need for such terms makes
one wonder to which extent is $\chi$PT-behaviour observed:
indeed, their r\^ole in the fits is to approximately cancel the curvature of the NLO
chiral logarithms and restore a roughly linear dependence in a wide range of $m_\pi^2$.
A possible reading of this finding is that the chiral logarithms
set in only beyond the pion masses well probed by our data.
Similar observations, namely that the NNLO terms effectively cancel out the NLO
logarithms, thus restoring a linear behaviour (e.g.~for $f_\pi$), and that linearity sets
in at relatively low pion masses already, have been reiterated in the detailed
analysis of \cite{Durr:2013goa}.
The continuum and chiral limits are carefully taken thanks to our ensembles
covering a comfortable region in the $(a,m_\pi)$ plane (in particular,
the lightest-pion ensemble G8 seems pivotal for stability).

To complete this preliminary investigation, a similar work on the
$\xi$-expansion for $\chi$PT formulae (i.e.~in terms of quark mass
instead of pion mass squared), as well as full inclusion
of NNLO chiral logarithms, are planned, with the goal
of a better assessment of systematic uncertainties; moreover,
comparative analyses of the same data, based on the
Sommer scale $r_0$ \cite{Sommer:1993ce}
and on the recently introduced flow-time
scale $t_0$ \cite{Luscher:2010iy},
are currently being undertaken \cite{Bruno:pos_lat_13}.
A more complete and articulated account 
of the present analysis
is deferred to a forthcoming paper.

The authors gratefully acknowledge access to HPC resources in the form of
a regular GCS/NIC project$^1$, a JUROPA/NIC project\footnote{See
\href{http://www.fz-juelich.de/ias/jsc/EN/Expertise/Supercomputers/ComputingTime/Acknowledgements.html}
{\tiny{\texttt{http://www.fz-juelich.de/ias/jsc/EN/Expertise/Supercomputers/ComputingTime/Acknowledgements.html}}}.}
and through PRACE-2IP, receiving funding from the
European Community's Seventh Framework Programme (FP7/2007-2013) under grant
agreement RI-283493.
This work is supported in part by the grants SFB/TR9
of the Deutsche Forschungsgemeinschaft.

\bibliography{lottini_lat2013_proceeding}
\bibliographystyle{JHEP}

\end{document}